\documentclass[11pt]{article}
\usepackage{graphicx}

\begin{document}

\title{ Aspects of the stochastic Burgers equation and their connection
with
turbulence.}
\author{F. Hayot and C. Jayaprakash \\ Department of Physics \\ The Ohio
State University \\ Columbus, Ohio 43210}

\maketitle

\vspace{1cm}

{\it Summary.}

We present results for the 1 dimensional stochastically
forced Burgers equation when the spatial range of the forcing varies.
As the range of forcing moves from small scales to large scales, the system
goes from a chaotic, structureless state to a structured state dominated
by shocks. This transition takes place through an intermediate region
where the system exhibits rich multifractal behavior. This is mainly the
region of interest to us. We only mention in passing
the hydrodynamic limit of
 forcing confined to large scales, where much work has taken place
since that of Polyakov\cite{pol}.

In order to make the general framework clear, we give an introduction to 
aspects of isotropic, homogeneous turbulence, a description of
Kolmogorov
scaling, and, with the help of a simple model, an introduction to the
language of multifractality which is used to discuss intermittency
corrections to scaling.

We continue with a general discussion of the Burgers equation and forcing,
and some aspects of three dimensional turbulence where - because of the
mathematical analogy between equations derived from the Navier-Stokes and
Burgers equations - one can gain insight from the study of the simpler
stochastic
Burgers equation. These aspects concern the connection of dissipation rate
intermittency exponents with those characterizing the structure functions
of the velocity field, and the dynamical behavior, characterized by
different time constants, of velocity structure functions. We also
show how the exponents characterizing the multifractal behavior of velocity
structure functions in the above 
mentioned
transition region can effectively be calculated in the case of the
stochastic Burgers equation.

\vspace{2cm}

{\bf Table of contents.}

\vspace{0.5cm}

I. {\bf Introduction.}\\
I.1. Kolmogorov scaling.\\
I.2. A simple model.\\
I.3. The language of multifractality.

\vspace{0.5cm}

II. {\bf The stochastic Burgers equation.}\\
II.1. Shock structure and extreme multifractality.\\
II.2. Stochastic forcing.\\

\vspace{0.5cm}

III. {\bf Three dimensional turbulence and the stochastic Burgers
equation.}\\
III.1. Multifractal exponents.\\
III.2. Dissipation rate correlation and intermittency.\\
III.3. Dynamic behavior.

\vspace{0.5cm}

IV. {\bf Remarks on intermittency.}

\newpage

{\bf I. Introduction.}

We study some aspects of statistically stationary, homogeneous
and isotropic fully developed turbulence. This is the typical framework
in which such studies
are done. The quantities of interest are the equal time spatial
correlations of the velocity field $\vec u(\vec x,t)$, the so-called structure
functions. The longitudinal structure functions, which are the ones usually
discussed, are defined by
\begin{equation}
S_p(r)=<[(\vec u(\vec x + \vec r,t)-  \vec u(\vec x,t)).\vec n]^p> 
\end{equation}
where $\vec n$ is the unit vector in the direction $\vec r$. 
Some
components of the velocity field
difference can be projected onto the direction transverse to $\vec n$, and
thus there are other correlations of $p-th$ order, which involve
longitudinal 
and  an (even) number of transverse projections. \\
The velocity satisfies the incompressible Navier-Stokes equation
\begin{equation}
\partial_t \vec u +\vec u.\vec \nabla \vec u=  -\vec \nabla p 
+\nu \triangle \vec u+\vec f
\end {equation}
with 
\begin{equation}
\vec \nabla. \vec u=0
\end{equation}
Here $p$ is the pressure divided by the constant mass density, $\nu$ the
kinematic viscosity. We have added  $\vec f=\vec f(\vec x,t)$, an
external stochastic  
force which acts on large scales, and maintains a turbulent steady state.
The average in (1) then includes as well an average over time.\\
In the usual picture of turbulence (see I.1.), when the distance $r=|\vec r|$
in (1) is small compared to large scales $L$ of the order of the system size,
and large compared to the scales where dissipation takes place, the structure
functions are expected to behave as
\begin{equation}
S_p(r) \sim (r/L)^{\zeta_p}
\end{equation}
An important aspect of solving the problem of 
statistical isotropic, homogeneous turbulence
is deriving the values of the exponents $\zeta_p$ in (4) from
the Navier-Stokes equation. This has not been done except for $\zeta_3$,
the value of which is fixed by the Von Karman-Howarth relation\cite{monin}.
It turns out however that the experimentally measured $\zeta_p$'s
(up to $p=10$ or so) are not too different from their
scaling values as they arise in the picture of fully developed turbulence
proposed by Kolmogorov. This is the reason a
large
number of phenomenological models exist, which by breaking scale invariance
slightly, give improved fits to the data. The usual language in which to
express deviations from scaling is that of multiscaling or
multifractality. 

We will therefore discuss first in this introductory section
Kolmogorov scaling, then a simple model, which allows
one to introduce non-scaling elements, and provides a simple introduction to
the language of multiscaling which we present next. A general reference for
these subjects is the book of Frisch\cite{frisch}.

	In the second section we discuss the stochastic Burgers equation,
its shock structure and the associated extreme multifractality, and its
behavior 
when the spatial range of the random forcing varies from small to large
scales. In section III we take up the point about statistical aspects of the
stochastic Burgers equation and their connection with three dimensional,
forced,
isotropic and homogeneous turbulence. First we show how the problem of
multifractality can be solved for the stochastic Burgers equation. Then we
 discuss the relation between intermittency in the energy dissipation to
intermittency in the velocity field, and end up by making a number of
observations concerning the dynamical behavior of structure functions.
General remarks about intermittency in
fully developed turbulence and for the stochastic Burgers equation 
are made in section IV.

This report is based on a number of results or points made in
references\cite{f1,f2,f3,f4,f5}.

\vspace{0.5cm}

{\bf I.1. Kolmogorov scaling.}

\vspace{0.5cm}

The picture is that of an energy cascade from the large scale $L$ where
the energy is put into the system, to the dissipation scale $\delta$ where
it is dissipated. On intermediate scales $\delta \ll r \ll L$, which
make
up the
so-called
inertial range, the only quantity which matters is $\epsilon$, the
mean energy
dissipation
rate per unit mass, considered to be independent of scale.
 $\epsilon$ has the dimension of velocity squared
divided
by time, or velocity cubed divided by distance. 

The dissipation scale
$\delta$
can only depend on $\nu$ and $\epsilon$, and thus for dimensional
reasons $\delta\sim
(\nu^3/\epsilon)^{\frac{1}{4}}  \sim (1/Re)^{\frac{3}{4}}L$, where after
replacing $\epsilon$ in terms of a characteristic velocity $U$ and the
large scale $L$, we are able to introduce the Reynolds number $Re=UL/\nu$.
In the limit of small viscosity or large Reynolds number there is thus 
a definite separation of scales between $\delta$ and $L$.\\
In the inertial region, dimensions are determined by $\epsilon$ alone, and
therefore
one predicts
on dimensional grounds, that $S_p(r)$ which has the dimension of velocity to
the $p-th$ power behaves as
\begin{equation}
S_p(r) \sim \epsilon^{\frac{p}{3}} r^{\frac{p}{3}}
\end{equation}
This is Kolomogorov scaling. The scaling values of the exponents in (4)
are then
\begin{equation}
\zeta_p=p/3
\end{equation}
This gives $\zeta_2=2/3$, which by Fourier transform is equivalent
to the experimentally observed $-5/3$ behavior of the energy spectrum,
namely $E(k)=k^2<{\vec u}(\vec k). {\vec u}(-\vec k)> \sim
{\epsilon}^{\frac{2}{3}}
k^{-\frac{5}{3}}$. One also obtains $\zeta_3=1$, which is the value
fixed by the Von Karman-Howarth relation. The other general
result\cite{frisch} is
that $\zeta_p$ is a convex function of $p$. Measurements
of
the structure functions show\cite{antonia} however 
that Kolmogorov scaling does not hold:
the measured $\zeta_p$'s for $p>3$ lie below the scaling values. For
instance $\zeta_6=1.80\pm0.05$ rather than the scaling value of $2$, obtained
from (6).
This effect is called 
intermittency or multifractality, and can be
related heuristically to
the non-space filling property of the eddies which make up the energy
cascade, and therefore to their fractal dimension. A simple model will serve
to
illustrate these points.

\vspace{0.5cm}

{\bf I.2. A simple model.}

\vspace{0.5cm}

Among models which describe the energy cascade, the so-called $\beta$-model
\cite{sulem} is instructive.
Imagine, as the energy cascades down to smaller scales from the 
large scale $L$, that at scales $r=\alpha^nL$ in the inertial range, the 
eddies at this scale, which themselves have a typical size of $r$, occupy
only a fraction $\beta$ of the available
space, such that $p_r=\beta^n$, where $p_r$ can be interpreted as the
probability of finding an eddy of size $r$ at scale $r$. Eliminating the
"generation"
number $n$ between the expressions for $r$ and $p_r$, on finds
\begin{equation}
p_r = (r/L)^{3-D}
\end{equation}
where $3-D=\ln \beta/\ln \alpha$. If the eddies are space filling, then
$\beta=1$, and therefore $D=3$. The value of $3$ corresponds to the
fact that
we pretend our discussion is for eddies in 3 dimensions. The argument
itself is clearly independent of space dimension. One now interprets
$D$ as the fractal dimension of the space on which the eddies exist,
assuming that $D$ is smaller than $3$. 

What are the structure functions in this model? 

The typical energy of an eddy of size $r$ is $E_r\sim {\delta v_r}^2 p_r$,
and therefore the average energy dissipation rate (per unit mass) at
scale $r$, with a typical time scale $t_r=r/\delta v_r$, is
\begin{equation} 
\epsilon_r \sim  \frac{{\delta v_r}^3}{L} (r/L)^{3-D-1}
\end{equation}
Here $\delta v_r$ is the velocity variation across the eddy.
The value of $\epsilon_r$ is independent of $r$ if homogeneity holds (existence
of an inertial scale),
and therefore one has for the velocity
\begin{equation}
\delta v_r \sim (\epsilon L)^{1/3}(r/L)^{\frac{1}{3}-(3-D)/3}
\end{equation}
from which follows for the structure function
\begin{equation}
S_p(r)=<{\delta v_r}^p>= \delta v_r^p p_r \sim (\epsilon L)^{p/3}
(r/L)^{p/3+(3-D)(1-p/3)}
\end{equation}
One thus finds for the exponents $\zeta_p$ of the structure functions,
a convex function of $p$, namely
\begin{equation}
\zeta_p=p/3+(3-D)(1-p/3)
\end{equation}
which satisfies the condition (Von Karman-Howarth relation) $\zeta_3=1$.
The scaling violating part in $\zeta_p$ is given by 
 $(3-D)(1-p/3)$. 
 For instance $\zeta_6=2-(3-D)$, which, by comparison with the 
experimental result $\zeta_6=2-0.2$, leads to a fractal dimension $D=2.8$.
Note that the velocity variation at $r$ ($\delta v_r \sim r^h$) is itself
characterized by
an
exponent $h=1/3-(3-D)/3$. For $D=3$, when the eddies fill all space at any
inertial scale, one has the scaling (Kolmogorov) result $h=1/3$ and
$\zeta_p=p/3$.

In the simple model we have considered, the structure functions 
and the variations
of the velocity field are characterized by a single $h$ and $D$.
However here, as opposed to the Kolmogorov scaling
behavior, the
eddies are not space filling, but are characterized by a fractal dimension
$D$. 

Simple fractal models such as the one we have described are not believed to
give the whole picture required to describe fully developed turbulence.
Experimental data suggest that $\zeta_p$ depends non linearly on $p$ in
contrast to equation (11).
It is believed\cite{frisch} that one needs to consider a more general picture,
with a range of
possible $h$'s and of corresponding fractal 
dimensions $D(h)$ (see section IV.).
 This picture, or the language in which it is formulated,
is that of multifractality, which we discuss next.

\vspace{0.5cm}

{\bf I.3. The language of multifractality.}

\vspace{0.5cm} 

Assume now that $h$ can take values in an interval
$(h_{min},h_{max})$, and that to each $h$ there corresponds a set in three
dimensional space of fractal dimension $D(h)$, in such a way that across
any
distance $r$ ( $r$ belongs to the inertial range) in the vicinity of that
set, one has
\begin{equation}
\delta v_r \sim (r/L)^h
\end{equation}
and
\begin{equation}
p_r \sim (r/L)^{3-D(h)}
\end{equation}
where $p_r$ is the probability for being within a distance of the set of
fractal dimension $D(h)$, and $\delta v_r$ is the velocity variation.
As a consequence one has the following expression for $S_p(r)$
for a given set with scaling dimension $h$
\begin{equation}
S_p(r) \sim <{\delta v_r}^p> \sim (r/L)^{ph+3-D(h)}
\end{equation}
All $h$ can contribute to the right-hand side, but since $r/L \ll 1$, the
dominant exponent $\zeta_p$ is given by
\begin{equation}
\zeta_p= \min_{ h}(ph+3-D(h))
\end{equation}

This exponent $\zeta_p$ is the dominant one in the expression of the structure
factors (cf. equation (4)).

{\bf Remarks:}

- the scaling result corresponds to $h=1/3$ and $D(1/3)=3$.\\
- the argument is the same in 1 or 2 dimensions with the replacement of
the number $3$ in $3-D(h)$ by respectively $1$ and $2$.\\
- the quantity $3-D(h)$ is positive or
zero, since $D(h)$ cannot exceed the
dimension of the embedding space. It is generally assumed that $h_{min} \geq 0$.
In the case of the Burgers equation where exponents can be calculated, we find
(cf. section III.1.) that the $h$'s corresponding to higher order structure
functions reach the value 0 when the stochastic forcing has moved to
sufficiently large scales, and stay at the value 0 when the scale of the
forcing increases further.  

\vspace{0.5cm}

{\bf II. The stochastic Burgers equation.}

\vspace{0.5cm}

This is a 1 dimensional version of the Navier-Stokes equation, a version 
without 
incompressibility and
pressure, which describes the evolution of the compressible
field $u(x,t)$, by
\begin{equation}
\partial_t u +u\frac{\partial u}{\partial x}= \nu \frac{\partial^{2}u}{
\partial x^2} + f
\end{equation}
where $f=f(x,t)$ is a stochastic forcing.

We will discuss later the forcing and its influence on the dynamics
of the field. For the moment, we will ignore it, and summarize some
results concerning the plain Burgers equation\cite{burgers}. 

\vspace{0.5cm}

{\bf II.1. Shock structure and extreme multifractality.}

\vspace{0.5cm}

If one starts from an initial sinusoidal velocity profile 
of large wavelength, then 
under the influence of the nonlinear term in the equation, the sinusoid
will for sufficiently small viscosity,
steepen into a series of shocks. After some time the shocks will fade
away, 
their energy being dissipated by the viscous term. This viscous term plays
a role mainly at the position of the shocks, where it is counterbalanced by
the nonlinear term. The equality of these two terms leads to
\begin{equation}
\nu=\triangle u .\delta
\end{equation}
where $\triangle u$ is the velocity jump across the shock, and 
$\delta$ is the shock
thickness. There are thus two scales here: a large scale $L$ corresponding
to some average distance between shocks, and a dissipation scale $\delta
\sim \nu$, very much smaller than $L$ when $\nu$ goes to zero. Distances
away from both extremes make up the inertial range.

In terms of multifractal language, 
the Burgers equation (
one averages, in the limit $\nu \rightarrow 0$, over an ensemble of
initial states, or considers stochastic forcing on large scales) shows extreme
multifractality, a situation called bifractality in the 
literature\cite{frisch}. 
The behavior of $u$ is essentially linear 
between shocks ($u \sim x$), and thus here $h=1, D(1)=1$. At the shocks
themselves $h=0, D(0)=0$, since the shocks are discontinuities of the
velocity field occurring at a point (in the $\nu \rightarrow 0$ limit).
The velocity variation across the shock is independent of distance,
and the probability of being within a distance $r$ is linear in $r$ (cf. 
equations (12) and (13) for the case of 1 dimension).\\
There are thus two possible values for the exponent $ph+1-D(h)$ (cf.
section I.3.), namely $p$ or $1$, and therefore 
the dominant exponent $\zeta_p$ (equation (15))
characterizing
the behavior of the structure functions in the inertial scale, is such
that
\begin{eqnarray}
\zeta_p=1, & p \ge 1
\end{eqnarray}\
This is an extreme case of multifractality ( all exponents have
the same value for integer $p$ greater than 1), very much
different from the 
case of three dimensional homogeneous, isotropic turbulence where the 
experimentally determined 
 exponents remain relatively close to the scaling ones, which increase linearly
with $p$ (see equation (6)). 

However -  as we have discovered -  there is a whole range of
multifractal behavior as 
the spatial extent of the stochastic force in the Burgers equation 
varies, and the
situation is much more interesting.

\vspace{0.5cm}

{\bf II.2. Stochastic forcing.}

\vspace{0.5cm}

For the stochastic forcing in (16) we take a Gaussian, such that in
$k$ space
\begin{displaymath}
<f(k,t)> =  0  
\end{displaymath}
\begin{equation}
<f(k,t)f(k',t')>= 2 D_0|k|^{\beta} {\delta_{k,-k'}}\delta(t-t')
\end{equation}
The exponent $\beta$ determines over which scales the forcing acts. For
$\beta >0$ it acts effectively on small scales, whereas as $\beta$ becomes
negative, larger and larger scales matter. The limit relevant to forcing
in three
dimensional turbulence is that of large scales, of the order of the system
size $L$.

The range of values of $\beta$ goes from $\beta=2$, which corresponds to
thermal
noise, to $\beta=-3/2$. For values smaller than the latter, the 
statistics of the velocity field is independent of $\beta$,
unchanged
from its behavior at $\beta=-3/2$. At $\beta=-3/2$ the system behaves as
the steady state of the plain Burgers equation: it exhibits the 
extreme multifractal behavior discussed in II.1., characteristic 
of a shock dominated velocity field. For $\beta>0$ however, the presence of
noise on small scales prevents the shocks from developing, and therefore
the behavior appears chaotic, i.e. random and structureless. Thus 
as $\beta$ moves from positive to
large negative values, the velocity field goes from a
chaotic to a shock dominated state, through 
an intermediate
region\cite{weichman} 
($-3/2<\beta<0$), where for $-1<\beta<0$ it displays complex dynamics
of appearing, interacting and disappearing 
shocks. This region is one of rich multifractal behavior, and
is the principal object of our study. It is through this region that
one approaches the hydrodynamic limit of large scale forcing from a
purely chaotic state.

To be complete, we mention that for positive values of $\beta$ one
can use a renormalization group approach. As soon as $\beta$ becomes negative, 
all sorts of non-linear
terms become important in the equations, and the perturbative 
renormalization group
approach breaks down.
This approach has been usually 
applied\cite{kardar} to the equivalent 
KPZ (Kardar-Parisi-Zhang) equation
for
fluctuations of
an interface height $h(x,t)$, related to $u$ by $u=\partial_x h$.
With a noise of
the form considered, the renormalization group has also been applied 
to the Navier-Stokes
equation\cite{dedom}. 

For $\beta$ positive , close to zero, the scaling analysis 
leads to the following result for the exponents $z$ and $\zeta_2$,
which appear in the scaling form assumed for $S_2(r,\tau)=
<(u(x+r,t+\tau)-u(x,t))^2>$, namely $S_2(r,\tau)= r^{\zeta_2}g(\tau/r^z)$:
\begin{equation} 
z+\zeta_2/2=1
\end{equation}
and
\begin{equation}
\zeta_2-z=-1-\beta
\end{equation}
The first relation is a consequence of Galilean invariance, the second of
the fact that the coefficient $D_0$ of noise fluctuations is not
rescaled because of the non-analytic form of the noise. One obtains from (20)
and (21) that $\zeta_2=-2\beta/3$ and $z=1+\beta/3$.

We will from now on consider the region of negative $\beta$, which is
so to speak the gateway to hydrodynamic behavior.

\vspace{0.5cm}

{\bf III. Three dimensional turbulence and the stochastic Burgers equation.}

\vspace{0.5cm}

We believe that because of the mathematical similarity of the Navier-Stokes
equation
with forcing, and the stochastic Burgers equation, the latter can be used 
as a key to the understanding of a number of issues in the statistical behavior
of isotropic, homogeneous turbulence. In the work we have been
doing\cite{f1,f2,f3,f4,f5}, we highlight this
similarity on
a number of occasions, in different situations. To give a simple 
example here, we 
compare 
the Von Karman-Howarth relation for $S_3$ for both equations.

For the Navier-Stokes equation with forcing $\vec f$, this relation
takes the following form
for the (equal time) 3rd order structure 
function $S_{3j}=<(\vec u_1-\vec u_2)^2 (
u_{1j}-u_{2j})>$, where "1" refers to the point $\vec x+\vec r$, "2" to 
the point
$\vec x$, and "j" denotes the j-th component of $\vec u$
\begin{equation}
\frac{1}{2} \partial _{r_j} S_{3j}(r) =  \nu \triangle S_2(r)-2 <\epsilon>
+<(\vec
u_1-\vec u_2).(\vec
f_1-\vec f_2)> 
\end{equation}
where $S_2(r)=<(\vec u_1-\vec u_2)^2>$,
while for the stochastic Burgers equation, where $S_3(r)= <(u_1-u_2)^3>$, 
it reads
\begin{equation}
\frac{1}{6} dS_3(r)/dr= \nu d^2S_2/dr^2- 2 <\epsilon>+<(u_1-u_2)(f_1-f_2)>
\end{equation}
The structural similarity of the two equations is clear.
 
One can derive the above two
 Von Karman-Howarth relations in a straightforward way from the space
and time dependent $S_2$,  
using the homogeneity in time of expectation values. More precisely, one
writes that $\partial S_2(r,\tau)/\partial t_1 + \partial S_2(r,
\tau)/\partial t_2 = 0$, where $r= x_1-x_2, \tau=t_1-t_2$. This
derivation
highlights the fact, which we have several times pointed out in our work,
that it is often useful for deriving equal time correlations to pass
through time dependent calculations. Many identities can be obtained 
this
way.

The two equations (22) and (23) are very similar. The 3
dimensional
result contains Kolmogorov's "4/5th" law
for the longitudinal structure function. In
both cases $<\epsilon>$ represents the 
energy dissipation rate. Since $r$ belongs to the inertial scale 
the term multiplied
by $\nu$ is negligible in both equations in the zero viscosity limit. The noise
dependent term can be evaluated
in the equal time limit with the help of the Novikov-Donsker
formalism\cite{novikov}. When 
the noise
is cut-off at large scales (the hydrodynamic limit) this term
leads to a
subdominant correction of order $(r/L)^2$. We will discuss later, for the stochastic
Burgers equation,
the general case when the noise ranges over small scales as well.

Though this comparison of the Von Karman-Howarth relations is based on 
a simple case,
we have found that the same similarity term by term, with an obvious display
of the 3 dimensional space indices, holds for any other equation we have 
derived involving 
velocity or dissipation rate correlations, with the exclusion of course of 
terms involving pressure.

We will discuss in the following three main points:

(i) first, we are going to face for the stochastic Burgers equation the 
problem of turbulence, namely calculate, for small $p$, in 
the multifractal region
($-1<\beta<0$) the exponents $\zeta_p$  characterizing the statistical
behavior of velocity structure functions,

(ii) second, we are going to give the general equation satisfied by
the
equal time correlation of the dissipation rate, and connect its
intermittent behavior, which exhibits a hierarchy of exponents, to the
intermittent behavior of the velocity structure
functions,

(ii) third, we investigate the dynamics of the second order structure
function, and show how - even in the absence of any average flow - 
$S_2$ satisfies a wave equation with characteristic velocity $\sqrt
{<u^2>}$. These dynamic considerations enable us to disentangle,
in our Eulerian
 framework, the intrinsic dynamical and
the kinetic, ballistic characteristic times which describe
the time evolution of flow structures.

\vspace{0.5cm}

{\bf III.1. Multifractal exponents.}

\vspace{0.5cm}

We are interested in the region where $-1<\beta<0$. Here also exists 
the possibility of scaling behavior,
in the same way as there is Kolmogorov scaling for three 
dimensional turbulence,
where the dimension of
$<\epsilon>$
or equivalently $D_0$, determines the dependence on distance of the $S_p$'s in
the inertial range. One thus has 
\begin{equation}
S_p(r) \sim (D_0/L)^{p/3} r^{-p\beta/3}
\end{equation}
which corresponds to $\zeta_p= - p\beta/3$ and $h=-\beta/3$. This 
is the value of $h$ in the scaling
regime. 
(Notice that at $\beta=-1$ the exponents are the same\cite{yakhot} as those of
Kolmogorov
scaling, equation (6).) 

This scaling regime is however dominant only in the region of $\beta$
negative close
to zero, and gives way to multifractal behavior as $\beta$ goes towards $-1$.
We are going to study this behavior directly on equations for the structure
functions derived
 from the stochastic Burgers equation. We proceed systematically
discussing
first $S_2$ and $S_3$, and then $S_4, S_5$ and general $S_p$.

(i) {\bf $S_2$ and $S_3$.}

One cannot derive directly from the stochastic Burgers ( or from
the Navier-Stokes equation in three dimensions) a closed equation for the
equal time structure function $S_2$. We therefore check numerically that
$S_2(r)$ behaves in the following way 
\begin{equation}
S_2(r) \sim  (r/L)^{-2\beta/3}
\end{equation}
for all $-3/2<\beta<0$. Precise numerical results, and therefore a precise
value of the exponent, can be obtained from evaluating 
the energy spectrum ($E(k) \sim
|k|^{-1+2\beta/3}$),  related to $S_2$ by Fourier
transform, rather than from $S_2$ itself ( Figure 1). $S_2 (r)$ 
thus scales, in the sense
that $\zeta_2=-2\beta/3$
has its scaling value (cf. equation (24)).

\begin{figure}[h]
\centering
\includegraphics[width=3.2in]{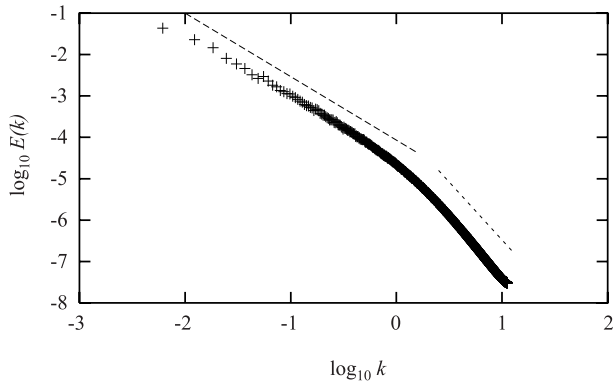}
\caption{ Graph of $\log E(k)$ as a function of $\log k$, where
the energy spectrum is $E(k) \sim |k|^{-1+2\beta/3}$,  
for $\beta=-0.8$. The straight line for small $k$, drawn for
comparison,  has a 
slope of $-1.53$, which is the value of $-1+2\beta/3$ at the given $\beta$.}
\end{figure}

As to $S_3(r)$, it is determined from the Von Karman-Howarth relation,
equation (23). In this equation the noise term takes in the equal time limit
(Novikov-Donsker formalism\cite{novikov}) the form
\begin{equation}
<(u_1-u_2)(f_1-f_2)>= 2 (1/L^2)  \sum _{k} D_0|k|^{\beta}(1-coskr)
\end{equation}
The term proportional to "1" in $(1-cos kr)$ cancels 
the $-2\epsilon$ in equation (23) because
$(1/L^2)\sum _{k} D_0 |k|^\beta$ is the total rate of energy input. One
thus obtains from equation (23) (in the $\nu \rightarrow 0$ limit)
\begin{equation}
\frac{1}{6} dS_3/dr = -2 (1/L^2) \sum_{k} D_0 |k|^{\beta} cos kr
\end{equation}
The "coskr" term leads by rescaling to the following result 
\begin{equation}
S_3(r) \sim  r^{-\beta}
\end{equation}
for $-1< \beta<0$,
in the case where the noise does not have a cut-off at scales of order $L$.
( At $\beta=-1$
there is an additional logarithm, $S_3 \sim r log r$.) 
 
The exponents characterizing the inertial range behavior of $S_2$ and 
$S_3$ have therefore
their scaling values throughout the domain $-1< \beta<0$. For
 $S_2$ the result is based on simulations, for $S_3$ the expression 
of the exponent
is obtained from the Von Karman-Howarth relation.

(ii) {\bf $S_4, S_5$ and general $S_p$.}

For $p\geq 4$ scaling no longer holds through the 
entire $-1<\beta<0$ range. The following are
the equations we obtain from the stochastic Burgers equation after isolating
the terms which in the inertial 
range go to zero when
the viscosity does, and simplifying the noise terms
\begin{equation} 
\frac{1}{6}dS_4(r)/dr= \frac{2}{3}\nu d^2S_3/dr^2-2<(\epsilon_1+\epsilon_2)
(u_1-u_2)>
\end{equation}
\begin{eqnarray}
\frac{1}{40}dS_5(r)/dr & = & \frac{1}{12}\nu d^2S_4/dr^2-\frac{1}{2L^2}\sum_{k}
D_0 |k|^\beta cos(kr)<(u_1-u_1)^2>  \nonumber \\
                       &   &  -\frac{1}{2}[<(\epsilon_1+\epsilon_2)(u_1-u_2)^2> \nonumber \\
                       &   & -<(\epsilon_1+\epsilon_2)> <(u_1-u_2)^2>]
\end{eqnarray}
\begin{eqnarray}
dS_p(r)/dr & \sim & \frac{1}{2L^2}\sum_{k}D_0 |k|^\beta
cos(kr)<(u_1-u_2)^{p-3}>
\nonumber \\
                  &      &  +...<(\epsilon_1+\epsilon_2)(u_1-u_2)^{p-3}>
\end{eqnarray}
The right-hand sides of equations (29) and (30) contain terms ( not
written for equation (31)) which go to zero in the small viscosity 
limit, a noise
dependent term and a dissipation rate dependent term.  The noise term 
has the general form
\begin{equation}
\sum _{k}D_0|k|^{\beta} cos (kr)<(u_1-u_2)^{p-3}> \sim \frac{dS_3(r)}{dr} 
S_{p-3}(r)
\end{equation}
since $dS_3(r)/dr \sim \sum_{k} D_0|k|^{\beta} cos kr$ (cf. equation (27)) 

Therefore scaling behavior in $S_p$ is present, whether dominant or
subdominant,
whenever there is scaling behavior in $S_{p-3}$. Thus the presence of a
scaling term in $S_2, S_3$ and $S_4$ guarantees the presence of one in any
$S_p$ for $p \geq 4$. We have already pointed out that both $S_2$ and
$S_3$ scale through the domain $-1< \beta <0$. The case of $S_4$ is
trickier because of the absence of an explicit noise term in equation
(29).  We discuss it below. First we turn to extracting the multifractal
behavior
of $S_4$ and higher order structure functions. This behavior becomes
relevant when the associated exponents are smaller than the scaling ones,
and therefore the corresponding non-scaling term dominates over the scaling
one, since $r/L \ll 1$.

We first note that in $k$-space both $S_3$ and $S_4$ depend on 
$<u(k_1)u(k_2)u(k_3)>, k_1+k_2+k_3=0$, the first one through its definition,
the
second one through the $\epsilon$ dependent term in (29). We thus make the
following general {\it ansatz}
\begin{equation}
Im <u(k_1)u(k_2)u(k_3)> \sim \frac {|k_1|^{\mu_1}|k_2|^{\mu_2}|k_3|^{\mu_3}}
{k_1k_2k_3}
+ permutations
\end{equation}
The constraint that $S_3(r) \sim r^{-\beta}$(cf. equation (28)) leads to
$\mu_1+\mu_2+\mu_3=
1+\beta$. We can show that the lowest exponent is obtained when 
$\mu_1=\mu_2=\mu_3= \mu/3 = (1+\beta)/3$. Putting the {\it ansatz} into the 2nd
term
of (29) leads to
\begin{equation}
dS_4/dr \sim \nu \int_{-\infty}^{\infty} d\alpha dk_1dk_2dk_3
\sin(k_1r)\frac{|k_1k_2k_3|^{\mu}}{k_1}\exp{-i\alpha(k_1+k_2+k_3)}
\end{equation}
Performing the $k$ integrals with a cutoff $\delta$ and then 
integrating over $\alpha$, with $0<\mu_1<1$, one obtains
\begin{equation}
dS_4/dr \sim \nu (2\pi/\delta)^{\mu_2+\mu_3}(1/\delta) r^{-\mu_1}
\end{equation}
and thus, with $\mu_1=\mu_2=\mu_3$,
\begin{equation}
S_4(r) \sim \frac{\nu}{\delta^{1+2\mu/3}} r^{1-\mu/3}
\end{equation}

It is important to note here that the non-scaling behavior arises from the
term in the equation which involves $\epsilon$.
The expression for $S_4$ contains two results:

(i) the fact that in the limit $\nu \rightarrow 0$ , 
\begin{equation}
\nu \sim \delta^{1+2(1+\beta)/3}
\end{equation}
whereas in the scaling limit $\nu \sim \delta^{1-\beta/3}$.
(By writing that at the dissipation scale $\delta$, the
characteristic eddy time $t_{\delta} \sim \delta/\delta^h$ is of order of
the
dissipation time ${\delta}^2/\nu$, one finds $\nu \sim \delta^{1+ h}$)
One thus has a new dissipation scale in $S_4$, namely
$\delta \sim \nu^{\frac{1}{1+h_4}}$.
This dissipation scale
depends 
on the corresponding multifractal
exponent $h_4=2(1+\beta)/3$. 
 For the dominant
term
this multifractal exponent has to be construed as the one
which minimizes $\zeta_p$ (cf. (15)).\\
(ii) second it gives the non-scaling exponent $\zeta_4=(2-\beta)/3$, which being
smaller than the scaling exponent $\zeta_4=-4\beta/3$ in the region $-1
< \beta < -2/3$, dominates over the scaling term in this region.

We now have to get back to the question how scaling behavior arises in $S_4$.
One can show that it arises through the 
$\nu dS_2/dr$ contribution in $S_3$ present in the Von Karman-Howarth
relation (cf. equation (23)). It corresponds to
$\mu_1+\mu_2+\mu_3=2+2\beta/3$ in the 
{\it ansatz} for $S_3$ (see above) with however $\mu_1 \neq \mu_2=\mu_3$.

One can now proceed along the same lines to find the behavior of $S_5(r)$,
taking as a starting point an {\it ansatz} similar to the one used for
$S_4$, but now for 
$<u(k_1)u(k_2)u(k_3)u(k_4)>, k_1+k_2+k_3+k_4=0$. There are now four $\mu$'s,
the sum of which is constrained by the known behavior of $S_4$ in two
different regions $-1 < \beta <-2/3$ and $-2/3 < \beta < 0$. We know already 
that in $S_5$ because of the presence in equation (30) of the noise term,
a scaling contribution will be present. The question that is to 
be settled through making
the {\it ansatz} on the 4-point function, is whether there are regions in which
the scaling term is subdominant, as happens for $S_4$. The answer is yes, and 
one  finds that there are three different regions:\\
(i) $-1/2 < \beta < 0$, where scaling behavior dominates, and thus
$\zeta_5=-5\beta/3$,\\
(ii) $-2/3 <\beta<-1/2$, where $S_5$ does not scale, $\zeta_5=(3-4\beta)/6$,
and this exponent is smaller than the scaling one and therefore the
corresponding term dominates in $S_5(r)$,\\
(iii) $-1<\beta<-2/3$, where $S_5$ has still another multifractal exponent,
$\zeta_5=(5-\beta)/6$, which gives the dominant behavior in this region of
$\beta$. The three exponents connect smoothly at the end points of each
interval. In each interval all three terms are present, but the term
with the smallest exponent dominates.
The first four $\zeta_p$'s are shown\cite{f3} in Figure 2.

\begin{figure}[h]
\centering
\includegraphics[width=3.2in]{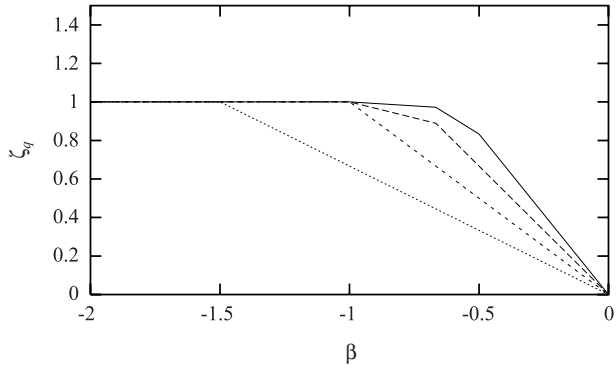}
\caption{ Exponents $\zeta_2, \zeta_3, \zeta_4$ and $\zeta_5$ vs. $\beta$,
for $-2<\beta<0$. The lowest curve is that of $\zeta_2$, the others 
above it are in order $\zeta_3, \zeta_4, \zeta_5$.} 
\end{figure} 

The following general
scenario emerges from these results: as $p$ increases, simple scaling with
$\zeta_p=-p\beta/3$
occurs over a progressively diminishing range of values for $\beta$ close
to
zero (and negative). Over most of the considered domain therefore,
multiscaling occurs as soon as $p \geq 4$, with the $\zeta_p$'s continuous
and  
piecewise linear, the number of linear segments increasing as $p$ gets
larger. As $\beta \rightarrow -1$ all the $\zeta_p$'s for $p \geq 3$
go towards $1$. This extreme multifractal regime is a manifestation of the
increasingly important role
played by shocks as the noise acts on larger and larger scales.

Several remarks are in order here:

(i) if one extracts a fractal scaling exponent for velocity variations 
from the calculations, as we have
done above for $S_4$ (equations (12) and (37)), one finds a
different value for $h_5$ in each of the three regions of $\beta$,
where different $\zeta_5$'s dominate, namely $h_5=-\beta/3$ for
$-1/2 < \beta <0$, $h_5=1/2+2\beta/3$ for $-2/3 < \beta <-1/2$,
and $h_5=(1+\beta)/6$ for $-1 < \beta < -2/3$. Thus $h_5$ is
continuous and piecewise linear, and goes to zero as $\beta \rightarrow -1$,
which is a reflection of the increasing dominance of
shocks. The same is
true for all $h_p$'s with $p \geq 4$.

(ii) one can also calculate continuous and piecewise linear fractal
dimensions $D(h_p)$ with the help of equation (15), assuming that the 
corresponding $h_p$
minimizes the right hand side, and using the values of $h_p$ and  $\zeta_p$
which result from the "{\it ansatz}" calculation. One finds that all fractal
dimensions tend towards zero as $\beta \rightarrow -1$, which again 
is consistent with the dominance 
of shock structure.

(ii) we cannot show in general that our calculation based on an
{\it ansatz} in $k$-space, and the assumption of the equality of $\mu$'s
in $S_5$
(cf. equations (33) and (34))
leads to the "true" dominant behavior in each domain. It is possible
that our continuous, piecewise linear $\zeta_p$'s, are only an
approximation to the "true" function $\zeta_p(\beta)$. 

\vspace{0.5cm}

{\bf III.2. Dissipation rate correlation and intermittency.}

\vspace{0.5cm}

By studying the full equation satisfied by the dissipation rate correlation
\begin{equation}
<\epsilon (x+r)\epsilon(x)> \sim <\epsilon>^2 (r/L)^{-\mu}
\end{equation}
we are able to find expressions for the intermittency exponent $\mu$
in terms of 
static and dynamic exponents of velocity field correlations.
Here $\epsilon(x)=\nu (\partial u/\partial x)^2$ for the Burgers equation and
$\epsilon(\vec{x})=\frac{\nu}{2}(\partial_i u_j+\partial_j u_i)^2$ for the
Navier-Stokes equation are the local dissipation rates. In our 
previous discussion, we have taken the
energy dissipation rate $\epsilon$ to be a constant, and this is all that is
required to obtain Kolmogorov scaling of the structure functions. In this
section $\epsilon(\vec x)$ is considered to be a fluctuating quantity which has
non trivial correlations, as experiment shows. One still has
$<\epsilon(\vec x)>
=\epsilon = constant$ because of homogeneity.

The following two relations have been proposed 
for the intermittency exponent $\mu$: 
\begin{equation}
\mu_1=2-\zeta_6
\end{equation}
and\cite{sulem, nelkin1}
\begin{equation}
\mu_2=2\zeta_2-\zeta_4
\end{equation}
The first one, the most discussed, because experimentally the value of $\zeta_6
\approx 1.8$
agrees with that of $\mu \approx 0.25$\cite{sreeni}, is essentially obtained
by a scaling
argument, which uses the dimension of $\epsilon$, namely $V^3/L$, to set
$<\epsilon(x+r)\epsilon (x)> \sim S_6(r)/r^2
\sim (r/L)^{\zeta_6-2}$.

The advantage of our approach lies in the fact that relations between $\mu$
and structure function exponents $\zeta_p$ are derived directly, and
simultaneously,
 from the 
equation satisfied by the dissipation rate correlation. This equation can
be derived from the stochastic Burgers or the Navier-Stokes equation by 
considering correlations in both space $r$ and time $\tau$, and then passing to
the $\tau \rightarrow 0$ limit. In this limit the noise term can be
expressed using the Novikov-Donsker formalism\cite{novikov}. One finds in this
way,
with $\epsilon_1=\epsilon(x+r,t+\tau), \epsilon_2=\epsilon(x,t)$
\begin{eqnarray}
<\epsilon_1\epsilon_2> & = & \frac{1}{4}
\partial_{\tau}<(\epsilon_1-\epsilon_2)
(u_1-u_2)^2> -\frac{1}{6} \partial_r<(\epsilon_1+\epsilon_2)(u_1-u_2)^3> 
\nonumber \\
                       &   & -\frac{1}{4}
\partial_r<(u_1-u_2)^2(\epsilon_2u_2-\epsilon_1u_1)> +\frac{\nu}{4}
{\partial}^2_r
<(\epsilon_1+\epsilon_2)(u_1-u_2)^2> \nonumber \\
                       &   & +\frac{1}{2} <(u_1-u_2)(\epsilon_2 f_1- \epsilon_1
f_2)>
\end{eqnarray}
The
3rd term on the right-hand side ensures Galilean invariance 
together with the first term (the left-hand side is Galilean invariant). The
viscosity dependent term, which is connected
to $d^3S_5/dr^3$ (cf. equation (30)), goes to zero for inertial
$r$
in the
zero viscosity limit. 

In order to show again the mathematical similarity of expressions derived
from the Burgers and Navier-Stokes equations, we show the equivalent
expression in three dimensions derived from equation (2):
\begin{eqnarray}
<\epsilon_1\epsilon_2> & = & \frac{1}{4}\partial_\tau
<(\epsilon_1-\epsilon_2)
(\vec u_1-\vec u_2)^2> 
                        -\frac{1}{4}\partial_{r_j}
<(\epsilon_1+\epsilon_2)(u_{1j}-u_{2j})
(\vec
u_1-\vec u_2)^2>  \nonumber  \\
                       &   & +\frac{1}{4} \partial_{r_j}<(\vec u_1-\vec
u_2)^2(\epsilon_1 u_{1j}-\epsilon_2 u_{2j})) \nonumber  \\
                       &   &  +\frac{\nu}{4}
{\partial}^2_ {r_j}<(\epsilon_1+\epsilon_2)(\vec u_1 - \vec u_2)^2>
                        +\frac{\nu}{2} \partial_{r_i}\partial_{r_j}
<\epsilon_1 u_{2i}u_{2j}+\epsilon_2 u_{1i}u_{1j}> \nonumber \\
                       &   & -\frac{1}{2} \partial_{r_i}<(u_{1i}-u_{2i})
(\epsilon_2 p_1+\epsilon_1 p_2)> \nonumber \\
                       &   & +\frac{1}{2}<(u_{1i}-u_{2i})(\epsilon_2
f_{1i}-\epsilon_1 f_{2i})>
\end{eqnarray}
Apart from the pressure term and a more complicated viscosity term due to
the difference in structure of the definitions of $\epsilon$ in the
Burgers and Navier-Stokes case (see the beginning of this section), the 
two equations correspond to each other
term by term, with an obvious generalization of space indices when going
from one to three dimensions. 

Now going back to equation (31) with $p=6$, one sees that the expression 
$<(\epsilon_1+\epsilon_2)(u_1-u_2)^3>$, which occurs in (41), is 
precisely the term in $dS_6/dr$
which, as argued in section III.1., leads to intermittency. Therefore from
(41), $<\epsilon_1\epsilon_2>$ (in the $\tau \rightarrow 0$ limit) contains
the intermittent behavior $(r/L)^{-\mu_1}$, with 
\begin{equation}
\mu_1=2-\zeta_6
\end{equation}
 as given in
equation (39).

As to the first term on the right hand side of (41), one can show\cite{f5} 
that 
the expression $<(\epsilon_1-\epsilon_2)(u_1-u_2)^2>$ appears in $\partial
S_4/\partial \tau$, where it is the only one involving the dissipation 
rate, and therefore leads to intermittency. There is thus a 
contribution here to the
intermittent behavior of $<\epsilon_1\epsilon_2>$ of exponent
\begin{equation}
\mu_2=z_{4,2} -\zeta_4
\end{equation}
where $z_{4,2}$ characterizes the behavior of the second order partial
derivative of $S_4$ in time, in the limit $\tau \rightarrow 0$.  The origin
of $\mu_2$ is thus dynamical. If
simple 
scaling in time holds, then $z_{4,2}= 2 z$, where $z=1-h$, with the value
of $h$ equal to its scaling value. 
$z$ here is the
dynamical exponent, not
the "frozen turbulence" exponent of value 1, which characterizes the
advection of small structures by large ones. Our preceding result and remarks
apply as well to Navier-Stokes turbulence. In this latter 
case $z=2/3$, which is
numerically equal to $\zeta_2$ (we are going to show in III.3. that this
result
is general and exact). Substituting $\zeta_2$ for $z$ (recall that in the
scaling limit $z_{4,2}=2z$) in (44) leads to the
result given in equation (40), which thus appears as 
a static approximation to what our
derivation shows to be the dynamical intermittency exponent given by 
equation (44).

For the Burgers equation the two intermittency exponents of equations (43)
and (44) are the two main ones. For the Navier-Stokes equation we can
only assert that these same two occur as well, because our discussion
does not take into account the pressure term in equation (42).

\vspace{0.5cm}

{\bf III.3. Dynamic behavior.}

\vspace{0.5cm}

Except for the discussion of $\mu_2$ in the preceding section, our 
concern up to now has been with the equal time structure functions.
We now address the problem of their dynamical behavior. We are interested
in relationships between dynamic and static exponents, and also in
shedding light on Taylor's frozen turbulence hypothesis in the case when 
there is no average flow field. In particular we wish to understand how
it happens that the square root of the rms fluctuations of the velocity
field replaces the average velocity when the latter is zero, thus
allowing ballistic behavior
 with $z=1$ ($z$ is defined by $\tau \sim r^z$). The objects of our study
are now the space and time dependent structure functions
\begin{equation}
S_p(r,\tau) = <(u_1-u_2)^p>
\end{equation}
where $u_1=u(x+r,t+\tau), u_2=u(x,t)$. The generalization to the three
dimensional case is straightforward.

We will concentrate on $S_2$. One can derive the following equation from the
stochastic Burgers equation\cite{f4}
\begin{equation}
\partial S_2(r,\tau)/\partial \tau = \frac{1}{2} \partial T_3/\partial r
+<u_1f_2> - <u_2f_1>
\end{equation}
where
\begin{equation}
T_3(r,\tau) = -<(u_1+u_2)(u_1-u_2)^2>
\end{equation}
which apart from additive constants is the same as $<u_1^2u_2+u_1u_2^2>$.
The term on the left-hand side and the first term on the right-hand side
form a Galilean invariant pair. In the $\tau \rightarrow 0$ limit $T_3$
does not contribute because of symmetry reasons.
In this limit there
is a
discontinuity in the noise term because $<u_1f_2>$ contributes for $\tau
> 0$, and $<u_2f_1>$ for $\tau < 0$. One thus has, using equations (23)
and (27),
\begin{equation}
\partial S_2(r,\tau=0^+)/\partial \tau = (1/L^2)\sum_{k}
D_0|k|^{\beta}coskr = -\frac{1}{12}dS_3/dr
\end{equation}
Assuming simple dynamic scaling for the first time derivative of 
$S_2$ (in the $\tau \rightarrow 0$ limit), with $\tau \sim r^{z}$, equation
(48) leads to the
following relation
\begin{equation}
\zeta_2-z= \zeta_3-1
\end{equation}
This equation is the same as equation (21). However here it follows
from an exact equation, whereas before it was obtained 
from a renormalisation analysis. Moreover $z$ here is precisely defined as the
exponent which characterizes the behavior
of the first order partial derivative of $S_2$ in time 
in the limit $\tau \rightarrow
0$.

Since $\zeta_3$ is known from the Von Karman-Howarth relation (equations
(22) or (23)), this equation relates the temporal and spatial exponents
which characterize the behavior of the 2nd order velocity structure
function. Since $\zeta_3$ has its scaling value set by the Von Karman-Howarth
relation, any scaling violations in
$\zeta_2$ has to be compensated by an equal one in $z$. 
Introducing the value of $\zeta_3$, one thus has in the case of the
Burgers equation
\begin{equation}
\zeta_2-z=-\beta - 1
\end{equation}
and in the case of Navier-Stokes
\begin{equation}
\zeta_2- z=0
\end{equation}
Thus $\zeta_2$ and $z$ are not independent, the knowledge of one
determines the other. This is the first constraint we have found for
$\zeta_2$, for which none can be found when one limits one's
investigations to static quantities only. In particular, 
in the Navier-Stokes case $\zeta_2=2/3=z$, whereas in the Burgers 
case one obtains
$z=1+\beta/3$. The latter results are consistent with the
simple Kolmogorov type scaling argument which entails $z=1-h$.

As $\tau \rightarrow 0$ what matters is clearly this dynamical $z$, the one
appropriate for a Galilean invariant situation. However as soon as 
$\tau$ departs from zero, the ballistic behavior with $z=1$ asserts itself. 
We have checked this numerically for $S_2(r=0,\tau)$ and $S_4(r=0,\tau)$,
for which, if dynamical scaling holds and for example $S_2(r,\tau)=
r^{\zeta_2} g(\tau/r^z)$, time dependence is of the form $\tau^{\zeta_2/z}$,
and similarly for $S_4$. Numerically one is able to distinguish\cite{f4} 
satisfactorily between the dynamic and ballistic values of $z$ (Figure 3). 
One thus verifies that as soon as $\tau$ is positive, ballistic behavior with
$z=1$ occurs.

\begin{figure}[ht]
\centering
\includegraphics[width=3.2in]{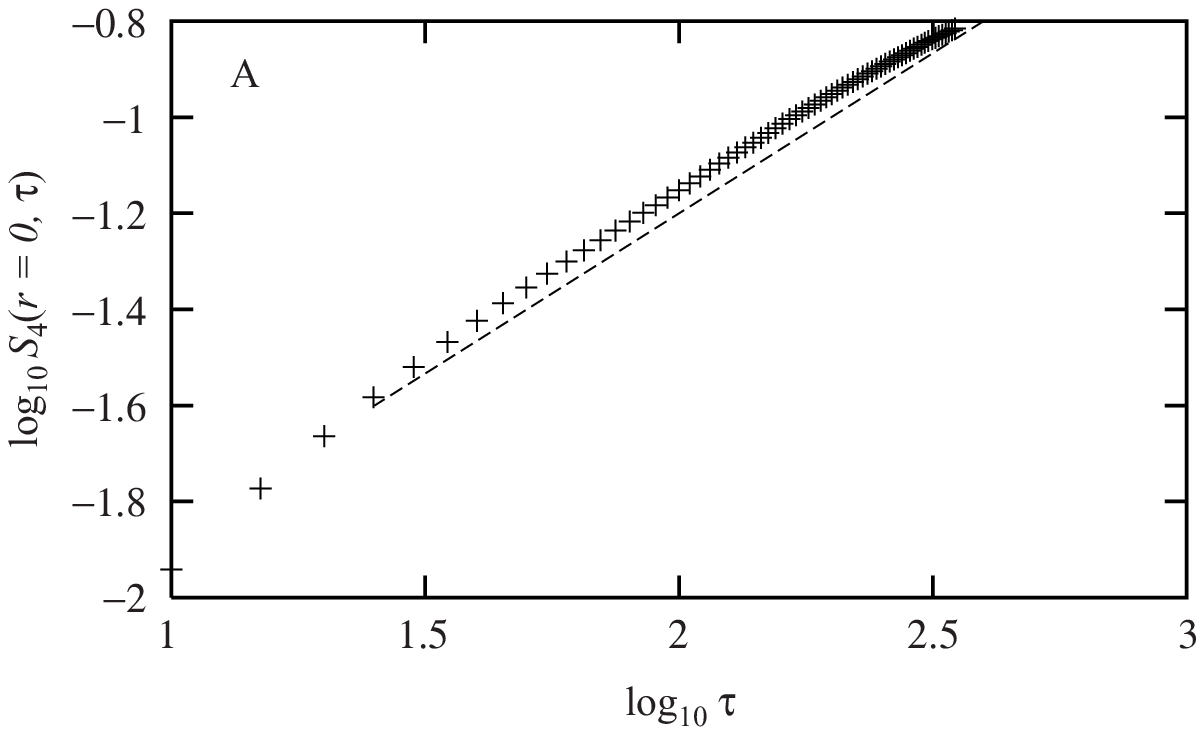}\\
\includegraphics[width=3.2in]{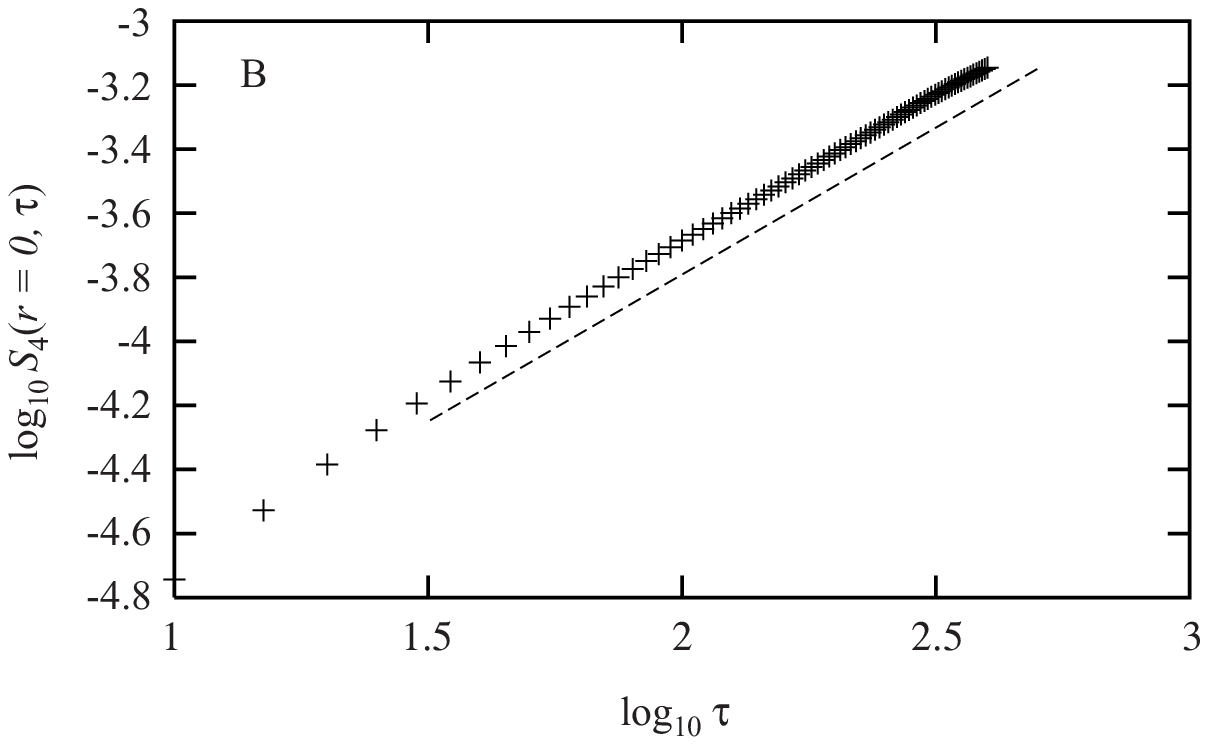}
\caption{ $Log_{10} S_4(r=0,\tau)$ vs. $log_{10} \tau$ for 
a) $\beta=-0.5$ with a dashed line of slope $\zeta_4/z=2/3$ with $z=1$,
and b) $\beta=-1$ where
the dashed line has a slope of 0.92 close to the numerically observed value of
$\zeta_4$. The expected slope is $\zeta_4/z$, and the numerical
results
allows one to distinguish between $z=1$ and $z=2/3$, the value of $z=1+\beta/3$
for $\beta=-1$.}
\end{figure}

The question now arises in which way ballistic behavior emerges, and
with it the use of Taylor's frozen turbulence hypothesis, in the
case when there is no average flow, i.e. $<u(x,t)>=0$. 

In reference \cite{f4} we have shown that if one differentiates
relative to
$\tau$
equation (46), one is lead to the following equation
\begin{equation}
{\partial}^2 S_2(r,\tau)/\partial \tau^2 = <u^2> {\partial} ^2 S_2/\partial
r^2 + ......
\end{equation}
The term on the right-hand side is a result of the fact that
\begin{equation}
\partial T_3/\partial \tau \propto <u^2> \partial S_2/\partial r
\end{equation}
after use of the assumption that in the $\nu \rightarrow 0$ limit the term
$<(u_1-u_2)^2 ({u_1}^2+{u_2}^2)> \sim 2 <u^2> S_2$. The latter assumption 
arises from the observation already made by Tennekes\cite{tennekes} that
large scales
eddies advect inertial scale information past an Eulerian observer.
Here we show that this assumption is encapsulated in the fact that
$S_2(r,\tau)$ satisfies precisely a wave type equation of characteristic
velocity given by the rms fluctuations of the velocity field. One expects this
behavior to occur over time scales large compared to the dissipation time
and small compared to the turnover time of the large scale structures in the
system. A detailed discussion of the other terms occurring in the equation
can
be found in reference \cite{f4} . 

\vspace{1cm}

IV. {\bf Remarks on intermittency.}

\vspace{0.5cm}

Before embarking on these remarks one should point out that the nature of
turbulence is different for the Burgers and Navier-Stokes equations: for
example vortex stretching is believed to be an important ingredient in three
dimensional developed turbulence.

Intermittency - the non-scaling behavior of the structure functions 
in the inertial range - is a hallmark of three dimensional turbulence.
The language of multifractality is a convenient way to describe it.
What is the origin of intermittency in the statistical behavior of turbulence?
The answer is not clear, though intermittency has been connected 
to the presence of 
vortex filaments in the flow. In one experiment\cite{tabeling}, where 
the size of the filaments is
several times the dissipation scale, they are associated with events in the
velocity field where the velocity derivative has large jumps. This is of course
what happens across shocks, which play the role of coherent structures
in the one dimensional stochastic Burgers equation. Here one has a
clear connection between intermittency and the
presence of shocks, though we are unable to give a numerical measure
of the number and sizes of shocks. Typically the velocity variation across a
shock occurs on length scales of the order of the dissipation scale.
For $\beta$ negative close to zero, shocks are barely apparent in the velocity
profile, and the structure functions show scaling behavior. 
As $\beta$ approaches $-1$ the shocks play a larger and larger role, and
intermittency, the difference between the actual values of the $\zeta_p$'s
and their scaling values, increases correspondingly (for $p \geq 4$). 
At $\beta \leq -3/2$
the shocks are present in full, dominating the velocity profile, and
intermittency is extreme: all $\zeta_p$'s are equal to $1$. There is
thus an obvious link between the dynamics of shocks - the small scale coherent
structures - and intermittency. 

We provide two other insights:

- we connect - not by a self-similarity argument, but from the exact equation -
the values of the exponents measuring intermittency in the energy 
dissipation rate 
to those measuring intermittency in the velocity 
structure functions (see III.2.),

- we show that in the equations for the velocity structure functions 
the terms responsible for intermittent behavior are those which contain
the energy dissipation rate. Intermittent behavior at the inertial scale 
is thus a consequence of dynamics which occurs at dissipation scales (see
III.1.).

For the stochastic Burgers equation we are 
of course able to provide an extra bonus: namely, with the help of an {\it
{\it ansatz}}, we are able to calculate from the basic equations 
the low order structure function exponents as
$\beta$ varies. Such a calculation remains the "holy grail" for
statistical three dimensional turbulence.\cite{nelkin2}

\vspace{0.5cm}

{\bf Acknowledgments.} This review paper was written while F. H. was on
sabbatical
at the Courant Institute of Mathematical Sciences at New York University.
He wishes to thank Dave McLaughlin and Mike Shelley for their hospitality and
support. 

We are also grateful to Mark Nelkin for a number of comments 
and suggestions he made concerning the manuscript. We thank the Ohio
Supercomputer Center for continuing support.

 \end{document}